\documentclass[a4paper,11pt]{article}
\usepackage{jinstpub} % for details on the use of the package, please see the JINST-author-manual
\usepackage{lineno}
\notoc

\title{A demonstrator for a real-time AI-FPGA-based triggering system for sPHENIX at RHIC}

 \author[a,1]{J. Kvapil,\note{Corresponding author.}}
 \author[b]{G. Borca-Tasciuc,}
 \author[c]{H. Bossi,}
 \author[d]{K. Chen,}
 \author[d]{Y. Chen,}
 \author[c]{Y. Corrales Morales,}
 \author[a]{H. Da Costa,}
 \author[a]{C. Da Silva,}
 \author[c]{C. Dean,}
 \author[a]{J. Durham,}
 \author[e]{S. Fu,}
 \author[f]{C. Hao,}
 \author[c]{P. Harris,}
 \author[c]{O. Hen,}
 \author[c]{H. Jheng,}
 \author[c]{Y. Lee,}
 \author[f]{P. Li,}
 \author[a]{X. Li,}
 \author[a]{Y. Lin,}
 \author[a]{M. X. Liu,}
 \author[e]{A. Olvera,}
 \author[g]{M. L. Purschke,}
 \author[h]{M. Rigatti,}
 \author[c]{G. Roland,}
 \author[i]{J. Schambach,}
 \author[a]{Z. Shi,}
 \author[h]{N. Tran,}
 \author[j]{N. Wuerfel,}
 \author[f]{B. Xu,}
 \author[k]{D. Yu,}
 \author[f]{H. Zhang}

 \affiliation[a]{Los Alamos National Laboratory,\\Bikini Atoll Rd, Los Alamos, NM 87545, United States}
 \affiliation[b]{Rensselaer Polytechnic Institute,\\ 110 8th St, Troy, NY 12180, United States}
 \affiliation[c]{Massachusetts Institute of Technology,\\77 Massachusetts Ave, Cambridge, MA 02139, United States}
 \affiliation[d]{Central China Normal University,\\ No.152, Luoyu Rd, Wuhan 430079, China}
 \affiliation[e]{University of North Texas,\\1155 Union Cir, Denton, TX 76205, United States}
 \affiliation[f]{Georgia Institute of Technology,\\225 North Ave, Atlanta, GA 30332, United States}
 \affiliation[g]{Brookhaven National Laboratory,\\PO Box 5000 Upton, NY 11973, United States}
 \affiliation[h]{Fermilab,\\PO Box 500. Batavia IL 60510, United States}
 \affiliation[i]{Oak Ridge National Laboratory,\\P.O. Box 2008. Oak Ridge, TN 37831, United States}
 \affiliation[j]{University of Michigan,\\500 S State St, Ann Arbor, MI 481091, United States}
 \affiliation[k]{New Jersey Institute of Technology,\\323 Dr Martin Luther King Jr Blvd, Newark, NJ 07102, United States}

\emailAdd{jakub.kvapil@lanl.gov; jakub.kvapil@cern.ch}

\abstract{The RHIC interaction rate at sPHENIX will reach around 3 MHz in pp collisions and requires the detector readout to reject events by a factor of over 200 to fit the DAQ bandwidth of 15 kHz. Some critical measurements, such as heavy flavor production in pp collisions, often require the analysis of particles produced at low momentum. This prohibits adopting the traditional approach, where data rates are reduced through triggering on rare high momentum probes.  We explore a new approach based on real-time AI technology, adopt an FPGA-based implementation using a custom designed FELIX-712 board with the Xilinx Kintex Ultrascale FPGA, and deploy the system in the detector readout electronics loop for real-time trigger decision.}

\keywords{%Only keywords from JINST's keywords list please:\\ https://jinst.sissa.it/jinst/help/keywordsList.jsp
Detector control systems (detector and experiment monitoring and slow-control systems, architecture, hardware, algorithms, databases);
Trigger algorithms;
Trigger concepts and systems (hardware and software)}
\begin{document}
\maketitle
\thispagestyle{empty}
\flushbottom

%\clearpage
\section{Motivation} \setcounter{page}{1}
\label{sec:intro}
Realizing the science potential of modern nuclear physics (NP) experiments at colliders relies on the collection and processing of very large data that often exceeds the current DAQ bandwidth and budgetary limitation. 
We propose to develop real-time AI technologies, implemented in the detector readout electronics loop, that address these challenges for the next generation of NP experiments at RHIC and Electron-Ion Collider (EIC). First, we will deploy a demonstrator that is being developed under the current Fast-ML project \cite{osti2022Accelerator} for the pp running in the sPHENIX experiment in 2024, and then generalize our approach for applications in experiments at the EIC, using future generations of EIC detector technologies.

The triggered readout rate of sPHENIX is limited to 15 kHz due to the design of the calorimeter readout system, which places limitations on the overall data volume and the expected collision rate. In pp collisions, RHIC delivers collision rates of 3 MHz, limiting sPHENIX to collect less than 1\% of heavy-flavour (HF) events of the total pp (and p+Au) rate when using triggered readout. The extended Streaming readout (SRO) of the tracking detectors can further improve the statistics up to 10\% of the total luminosity. The goal of this project is to sample the remaining luminosity further enhancing the collected data samples. The aim is to deploy a future system on the EIC to identify the (non)interesting Deep-Inelastic-Scattering processed in the e+p/A collisions.

\section{sPHENIX detector}
\label{sec:sphenix}
\begin{figure}[htbp]
\centering
\includegraphics[width=.49\textwidth]{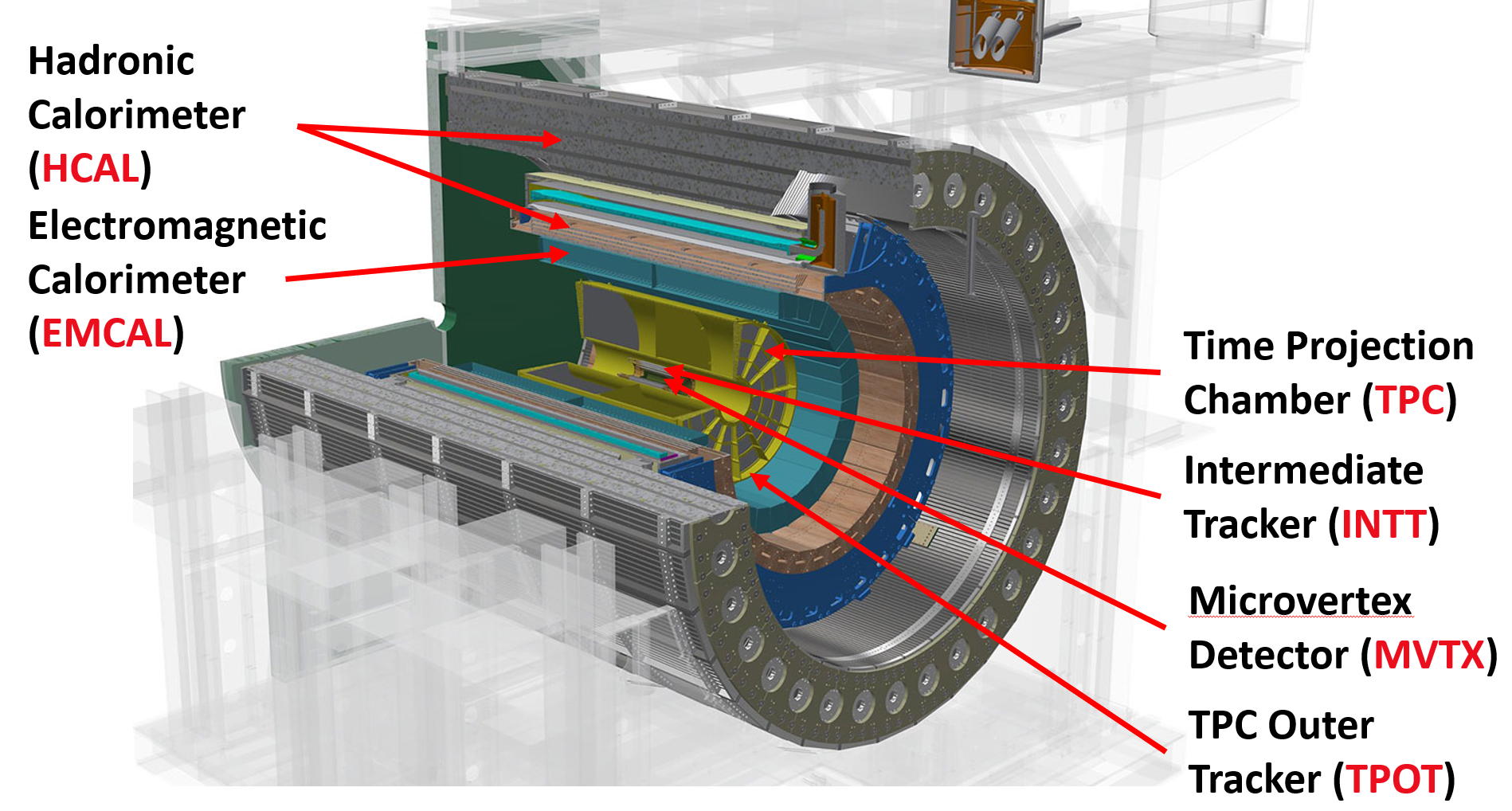}
\qquad
\includegraphics[width=.45\textwidth]{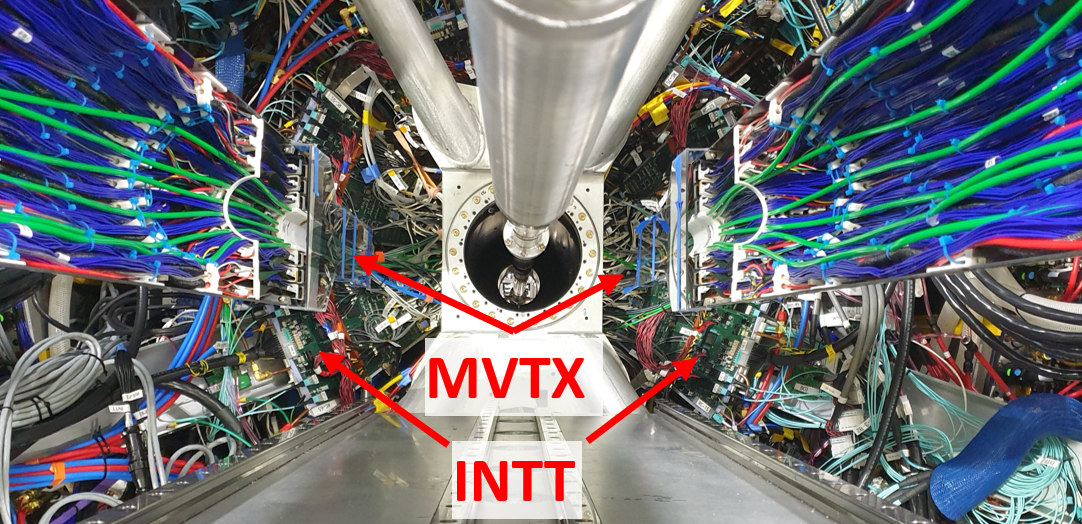}
\caption{Left: The sPHENIX detector with highlighted central barrel detectors. Right: Installation of MVTX and INTT detectors.\label{fig:1}}
\end{figure}

The sPHENIX detector \cite{cit:sphenix} is located at the RHIC accelerator complex in Brookhaven National Laboratory (BNL), with data taking in 2023 - 2025.The sPHENIX detector schematic and MVTX and INTT detectors are shown in figure~\ref{fig:1}. The tracking detectors (MVTX, INTT, TPC, TPOT) are capable of SRO. 
Even-though they are able to record all data, the data volume of the TPC detector exceeds the capability of the computing center and therefore a down-selection of what to save must be done. The aim of this project is to reconstruct tracklets from the silicon detectors in order to search for signatures of HF decays based on unique topology, and provide an additional trigger to sPHENIX.

\paragraph{Silicon detectors}
The three innermost tracking layers (the MVTX detector) are based on Monolithic Active Pixel Sensors (MAPS), the ALPIDE, originally developed for the ALICE experiment. The ALPIDE offers very fine pitch of 27 $\mathrm{\mu m}$ x 29 $\mathrm{\mu m}$ with a collision event time resolution of 5~$\mathrm{\mu s}$. The MVTX contain a total of 270M channels. The next two layers (the INTT detector) contain silicon strip sensors (manufactured by Hamamatsu) with a pitch of 78~$\mathrm{\mu m}$ x 16 (or 20) mm with 360k channels in total. Both detectors are shown in figure \ref{fig:1} (right).

\begin{figure}[htbp]
\centering
\includegraphics[width=.8\textwidth]{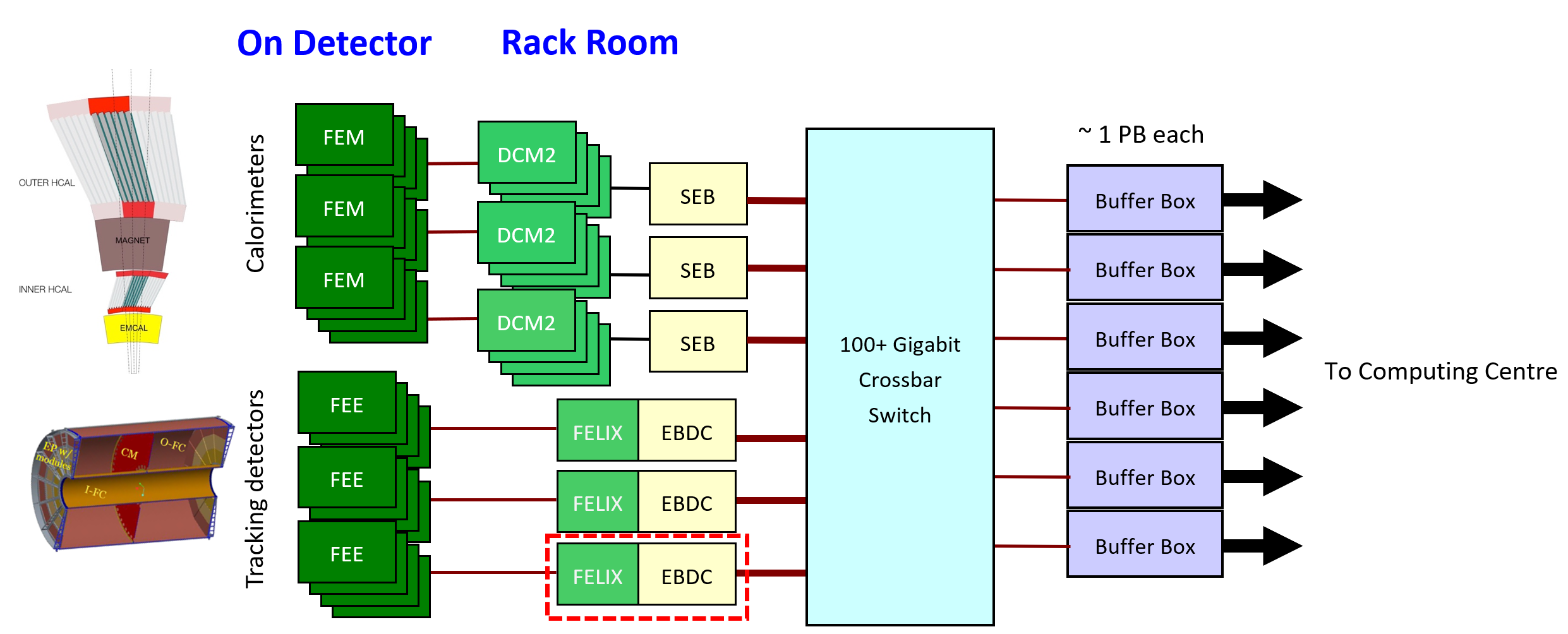}
\caption{A schematic of sPHENIX readout chain. The calorimetry data are sent from Front-End Modules (FEM) to the SubEvent Buffer (SEB) through the Data Collection Module (DCM2), and the tracking data from Front-End Electronics (FEE) are sent to the Event Buffer and Data Compressor (EBDC) through a FELIX interface card. \label{fig:2}}
\end{figure}

\paragraph{sPHENIX readout, trigger and timing distribution} 
The schematic of the sPHENIX readout chain is shown in figure~\ref{fig:2}. The tracking detectors' Front-End Electronics (FEE) sends data to the Event Buffer and Data Compressor (EBDC) through the FELIX (FELIX-712) \cite{cit:felix-712} interface card. The calorimetry detectors' Front-End Modules (FEM) send data to the SubEvent Buffer (SEB) through the Data Collection Module (DCM2). The trigger and timing information is generated by the Global-Level-1 trigger system (GL1). This information, in addition to detector-specific timing parameters and other book-keeping data, is adapted for each detector by Granule-Timing Modules (GTMs) that also transmit a copy of the accelerator clock to the front-ends, and the trigger information. The GL1 can currently generate up to 64 different triggers derived from 4 fiber inputs that carry the collision information from the front-ends, in addition to 4 hardware signal inputs that can be used for various purposes. The most used trigger inputs are from the Hadronic Calorimeters for the cosmic and high energy jet triggers (normally not pre-scaled) and Minimum-Bias Detector for the beam collision trigger (heavily pre-scaled). The goal is to provide an additional high efficiency trigger input for HF events in pp collisions.

\subsection{The DAQ-AI Data Flow}
\begin{figure}[htbp]
\centering
\includegraphics[width=0.8\textwidth]{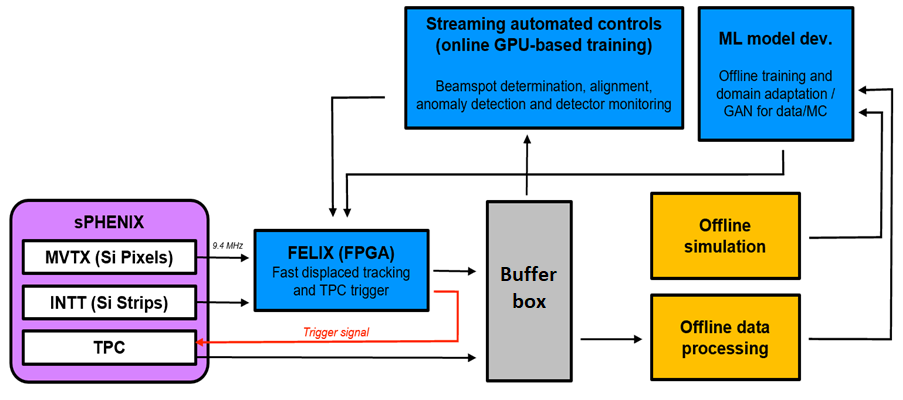}
\caption{The schematic of the DAQ-AI data flow.\label{fig:3}}
\end{figure}

Data from the MVTX and INTT are transmitted from the FELIX-DAQ board to the AI-Engine which is also based on FELIX board (FELIX-AI). It was decided that the AI-Engine will be hosted on a widely HEP community supported FELIX board for the following 3 reasons. First, FELIX offers 48 high-speed optical links to receive data; second, to reuse its Wupper module for the PCIe communication and associated software tools; and third, to use the sPHENIX infrastructure for tracking detectors that was built around the FELIX cards. The FELIX-AI houses the raw data decoder, event builder, clusterizer, and GNN models to provide fast tracking and the trigger decision. The trigger decision is sent via a LEMO cable to the GTM. Since the decision is based on an event topology, a reference point (beam spot), which changes in time, must be precisely known and monitored. A GPU based feed-back system will be in place to process the data from the buffer boxes, reconstruct the beam spot position and update the position in the FELIX-AI. The schematic of the data flow is shown in figure~\ref{fig:3}. 

There are 144 optical links running at 3.2~Gbps per fiber for the MVTX alone, thus two engines will be used, one for each MVTX/INTT hemisphere. This will allow to have 24 links for MVTX and 24 links for INTT. The FELIX optical links have been tested up to 14 Gbps with $\mathrm{BER}<10^{-16}$ with an external loop-back measurement. INTT offers excellent time resolution to tag 100~ns RHIC bunch-crossing time to assign a unique timing to each collision event.
\subsection{Latency breakdown} The TPC buffers can hold up to about 30~$\mathrm{\mu s}$ of data and that is the maximum hard limitation. Our stretched goal is to deliver the trigger decision to the gbloal system with in 10~$\mathrm{\mu s}$ which could allow us to capture data from other faster detectors. The MVTX contributes up to 5~$\mathrm{\mu s}$ latency, the cables between the interaction region and counting house contribute around 0.3~$\mathrm{\mu s}$ (81m fibres). Forwarding data to the FELIX-AI and decoding them takes up to 0.6~$\mathrm{\mu s}$, which results from the maximum depth of FIFO holding the decoded hits (128). This leaves around 4~$\mathrm{\mu s}$ for the AI engine to perform tracking and trigger decision. The real latency of the decoding and data transmission depends on the occupancy, which is estimated to be around 50 physics hits per chip per event. The FELIX-AI needs to ensure the latency is fixed, either by delaying the trigger decision, or vetoing the processing by decoding the bunch-crossing numbers from MVTX and INTT.

\section{Model description}

\begin{figure}[htbp]
\centering
\includegraphics[width=0.9\textwidth]{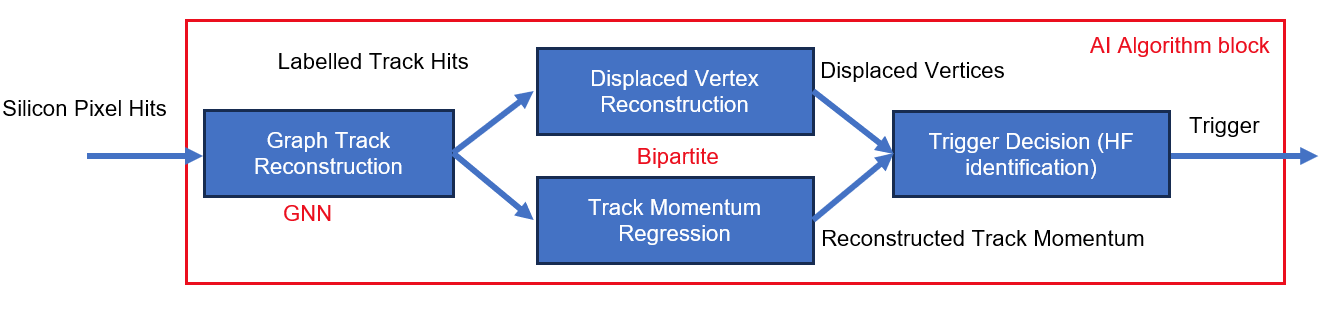}
\caption{Flowchart of the GNN network.\label{fig:4}}
\end{figure}

The model is based on Graph Neural Network (GNN) using PyTorch and PyTorch geometric. The aim is to reconstruct the decay topology of the tracklets and search for secondary (displaced) vertices, which is one of the prominent features of HF decays. The displaced vertex will be $\sim$ O(100~$\mathrm{\mu m}$) away from the primary vertex which has resolution $\sim$ O(10 $\mathrm{\mu m}$). We have developed a novel method for treating events as graphs consisting of tracks as nodes and interconnection between tracks when they belong to the same particle decay rather than hit graphs \cite{cit:model}. A 15\% improvement is observed by estimating the transverse momentum, $p_\mathrm{T}$, based on the tracks.  There are three stages in event processing: hit clustering, track reconstruction, and trigger decision. We use the GNN models to reconstruct tracks, remove outliers (TrackGNN), regress the track momentum and displaced vertices on the corresponding tracks, and to estimate the probability that the event is a trigger.  The flow chart is shown in figure~\ref{fig:4}. Table ~\ref{tab:i2} summarizes the efficiency and background rejection rates at two levels of signal/noise ratios, that is, 1\% and 0.1\%. The signal/noise ratio is computed using heavy flavor simulations where the $D^0$ decay products fall within the sPHENIX detector volume even if the tracks do not have sufficient momentum to reach all the tracking detectors. The test ratios can be compared to signal/noise ratio at sPHENIX under these circumstances which is expected to be approximately 0.5~\%. Note that the efficiency decreases with the signal/noise ratios. The apparent difference in efficiency is due to statistics of samples, however, the purity is decreased by a factor of 10 as expected. This is already comparable with sPHENIX partial streaming of 10\% MB events with $6.2\pm0.9$~\% purity for $D^0~p_T~\geq~2$~GeV~\cite{cit:sphenix-hf}.

\begin{table}
    \centering
\caption{Efficiency and Background Rejection with 1\% and 0.1\% signal/noise Ratios. \label{tab:i2}}
\smallskip
    \begin{tabular}{ccc}
    \hline
         \multicolumn{3}{c}{1\% signal/background ratio} \\
         \hline\hline
        Bg. rejection & Efficiency & Purity \\\hline
       90\% & 75.6\% & 7.1\% \\
        99\% & 19.4\% & 16.4\% \\
        \hline
    \end{tabular}
    \qquad
        \begin{tabular}{ccc}
    \hline
         \multicolumn{3}{c}{0.1\% signal/background ratio} \\
         \hline\hline
        Bg. rejection & Efficiency & Purity \\\hline
        90\% & 76\% & 0.75\%\\
        99\% &  23.2\% & 2.3\%\\
        \hline
    \end{tabular}

    \label{tab:my_label}
\end{table}

\section{Generation of the GNN IP core}

We propose to use High-Level Synthesis (HLS) tools to generate synthesizable Register Transfer Level (RTL) code, e.g., Verilog, to implement the software sPHENIX model onto FPGA. The steps are as follows. First, we manually translate the sPHENIX model into synthesizable C code and feed it into the HLS tool, Vitis HLS~\cite{cit:vitis-hls}. Then, we perform hardware optimizations of the model in HLS following the FlowGNN architecture~\cite{cit:flowgnn}, which is the state-of-the-art GNN architecture on FPGA. The target is to have a model implemented on FPGA that can process 100-200 nodes (hits) and 200-500 edges (hit connections) within  $\sim$ O(10 $\mathrm{\mu s}$).

\begin{figure}[htbp]
\centering
\includegraphics[width=\textwidth]{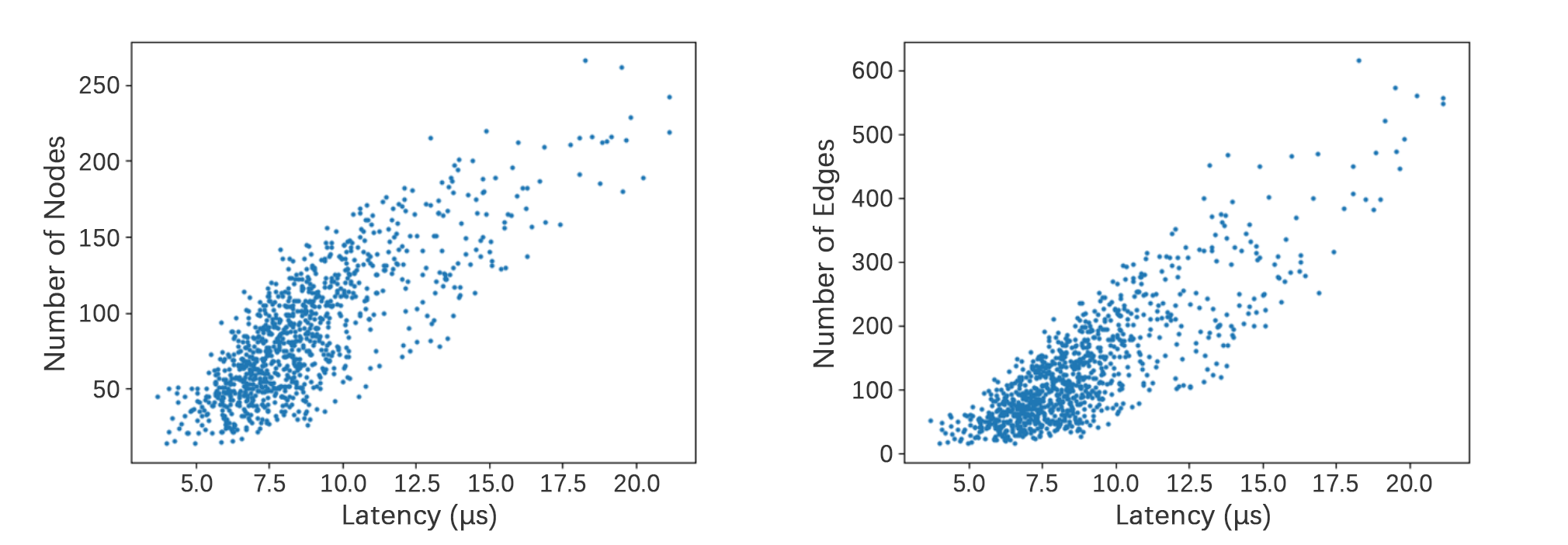}
\caption{The time-size scatter plot of the TrackGNN model measured at Alveo U280. Left: latency values with numbers of nodes. Right: latency values with numbers of edges. \label{fig:5}}
\end{figure}

\paragraph{FlowGNN Architecture with on-FPGA Implementation} 
The current TrackGNN model in sPHENIX we are using has one GNN layer, which includes 4 multi-layer perceptron (MLP) layers for both node and edge embedding with a dimension of 8.
The proposed architecture follows the message-passing framework in FlowGNN~\cite{cit:flowgnn}: the node embeddings are processed first, followed by an adapter to orchestrate the node information to the correct edge processing units for edge embedding computation and message aggregation.
We also use quantization to reduce the data precision and to reduce the memory and computation requirements. We use \texttt{ap\_fixed<18, 6>} for node embeddings, edge embeddings, and model weights and bias, and use \texttt{ap\_fixed<21, 9>} for messages and input node features. 
The initial study FPGA board is the Alveo U280, approximately twice as big as FELIX-712 used for the final deployment. The resource utilization is as follows: 194K (14.9\%) LUT, 214K (8.2\%) FF, 406 (20.2\%) BRAM, and 488 (5.4\%) DSP. The processing latency is measured on-board in an end-to-end fashion, including graph and weight loading, model computation, and results readback. The latency for an average-sized input (92 nodes and 142 edges) is 8.82 $\mathrm{\mu s}$ at a 285 MHz clock. The time-size scatter plot for nodes and edges are shown in figure \ref{fig:5}, where the x-axis is the time spent processing one graph, and the y-axis is the graph size in terms of the number of nodes and edges. Compared to CPU calculations, the FPGA was 99.86\% accurate.

\section{Summary and Outlook}
This paper describes how the streaming data of selected tracking detectors can be used to form an additional trigger input. Several modules of the FELIX-AI required for the trigger decision have been tested independently on different target devices. The final development is focused towards combining them into a single design and deploy on FELIX-712 boards. A second iteration of the TrackGNN model has achieved a latency of 8.82 $\mathrm{\mu s}$ at a 285 MHz clock, and this is sufficient for the target maximum latency limitation of 30~$\mathrm{\mu s}$. However it is approximately double from our stretched goal of around 4~$\mathrm{\mu s}$ in order to deliver the trigger decision within 10~$\mathrm{\mu s}$. A further improvement is expected by providing more lightened version of the model and by targeted optimisation of the design. Furthermore, a parallel effort to translate the model using the hsl4ml framework \cite{cit:hls}, which is a generalized package to translate neural networks into an IP core just started.

\acknowledgments
This project is funded by the United States Department of Energy, funding calls FOA-0002490, FOA-0002875 and Los Alamos National Laboratory LDRD program.

% Bibliography

%% [A] Recommended: using JHEP.bst file
%% \bibliographystyle{JHEP}
%% \bibliography{biblio.bib}

%% or
%% [B] Manual formatting (see below)
%% (i) We suggest to always provide author, title and journal data or doi:
%% in short all the informations that clearly identify a document.
%% (ii) please avoid comments such as "For a review'', "For some examples",
%% "and references therein" or move them in the text. In general, please leave only references in the bibliography and move all
%% accessory text in footnotes.
%% (iii) Also, please have only one work for each \bibitem.

\end{document}